\documentclass[]{emulateapj} 
\usepackage{latexsym}
\usepackage{mathrsfs}
\usepackage{times} 
\usepackage{graphics}
\usepackage[german,english,american]{babel}

\begin{document}

\shorttitle{The jet production efficiency of growing black holes}

\title{The kinetic luminosity function and the jet production
efficiency of growing black holes}

\author{S.~Heinz\altaffilmark{1}, A. Merloni\altaffilmark{2}, \&
J. Schwab\altaffilmark{3}} \altaffiltext{1}{Astronomy Department,
University of Wisconsin-Madison, Madison, WI 53706; {\em
heinzs\hbox{@}astro.wisc.edu}}\altaffiltext{2}{Max-Planck-Institute for
Astrophysics, 85743 Garching, Germany; {\em
am@MPA-Garching.MPG.DE}}\altaffiltext{3}{Massachusetts Institute of
Technology, MA 02139; {\em jschwab@mit.edu}}

\begin{abstract}
We derive the kinetic luminosity function for flat spectrum radio
jets, using the empirical and theoretical scaling relation between jet
power and radio core luminosity.  The normalization for this relation
is derived from a sample of flat spectrum cores in galaxy clusters
with jet-driven X-ray cavities.  The total integrated jet power at
$z=0$ is $W_{\rm tot} \approx 3\times 10^{40}\,{\rm
ergs\,s^{-1}\,Mpc^{-3}}$.  By integrating $W_{\rm tot}$ over
red-shift, we determine the total energy density deposited by jets as
$e_{\rm tot} \approx 2\times 10^{58}\,{\rm ergs\,Mpc^{-3}}$.  Both
$W_{\rm tot}$ and $e_{\rm tot}$ are dominated by {\em low} luminosity
sources.  Comparing $e_{\rm tot}$ to the local black hole mass density
$\rho_{\rm BH}$ gives an average jet production efficiency of
$\epsilon_{\rm jet}=e_{\rm jet}/\rho_{\rm BH}c^2 \approx 3\%$.  Since
black hole mass is accreted mainly during {\em high} luminosity
states, $\epsilon_{\rm jet}$ is likely much higher during low
luminosity states.
\end{abstract}

\section{Introduction}
\label{eq:introduction}
The $M-\sigma$ relation between black hole mass and the velocity
dispersion of the host galaxy's bulge \citep{gebhardt:00,ferrarese:00}
shows that the growth of black holes and large scale structure is
intimately linked.  X-ray observations of galaxy clusters show that
black holes deposit large amounts of energy into their environment in
response to radiative losses of the cluster gas
\citep[e.g.][]{birzan:04}. Finally, mechanical feedback from black
holes is believed to be responsible for halting star formation in
massive elliptical galaxies \citep{springel:05}.

These arguments hinge on the unknown efficiency $\epsilon_{\rm jet}$
with which growing black holes convert accreted rest mass into jet
power.  Constraints on $\epsilon_{\rm jet}$ are vital for all models
of black hole feedback.  Efficiencies of 1\% are typically assumed,
but this number is derived for phases of powerful jet outbursts only
and neglects the trend for black holes to become more radio loud at
lower accretion rates \citep{ho:01,gallo:03,merloni:03,falcke:04}.

The discovery of a tight correlation between core radio and X-ray
luminosity in accreting black hole X-ray binaries \citep{gallo:03} in
low luminosity states inspired a number of theoretical investigations
of how jet radio emission relates to the accretion state and rate of
the black hole. In the classical model by \cite{blandford:79}, the
flat spectrum radio synchrotron emission of a compact jet core is
produced by superposition of self-absorbed synchrotron spectra, each
from a different region in the jet. The model predicts a dependence of
the radio luminosity $L_{\nu}$ on jet power $W_{\rm jet}$ of the form
$L_{\nu} \propto W_{\rm jet}^{17/12}$.  More generally,
\cite{heinz:03a} showed that {\em any} scale invariant jet model
producing a powerlaw synchrotron spectrum with index $\alpha_{\nu}$
{\em must} obey the relation $L_{\nu} \propto W_{\rm
jet}^{(17+8\alpha_{\nu})/{12}}M^{-\alpha_{\nu}}$. With
$\alpha_{\nu}=0$ for flat spectrum jet cores, we can write
\begin{equation}
  W_{\rm jet}=W_0\left(\frac{L_{\nu}}{L_0}\right)^{12/17}
  \label{eq:radiopower}
\end{equation}
Given a measurement of $L_{\nu}$, we can thus estimate a jet's kinetic
power, up to a multiplicative constant $W_0$ (which we determine in
\S\ref{sec:w0}).  In \S\ref{sec:radio}, we will use this relation to
construct the kinetic jet luminosity function from the observed flat
spectrum radio luminosity function $\Phi_{L}(L_{\nu})$ (abbreviated as
FSLF below).  In \S\ref{sec:discussion}, we will derive the current
mean jet power per cubic Mpc and the jet production efficiency
$\epsilon_{\rm jet}$ of black holes.  Section \ref{sec:summary}
summarizes our results.  Throughout the paper we will use concordance
cosmological parameters of $\Omega_{\rm M}=0.3$,
$\Omega_{\Lambda}=0.7$, $H_{0}=75\,{\rm km\,s^{-1}\,Mpc^{-1}}$.

\section{The AGN-jet kinetic luminosity function}
\label{sec:radio}
As pointed out in \cite{merloni:04}\ and \cite{heinz:05b}, we can use
eq.~(\ref{eq:radiopower}) and the observed FSLF to derive the
underlying kinetic luminosity function $\Phi_{W}$ of flat spectrum
jets:
\begin{eqnarray}
  \Phi_{W}(W) & = & \Phi_{\rm
  L}\left(L_{\nu}(W)\right)\frac{dL_{\nu}}{dW} \nonumber \\ 
  & = & \Phi_{\rm L}\left({L_0(\frac{W}{W_0})^{17/12}}\right)
  \frac{17}{12}\frac{L_0}{W_0}\left(\frac{W}{W_0}\right)^{\frac{5}{12}}
  \label{eq:kineticlumi}
\end{eqnarray}
We will follow \citep[][DP90]{dunlop:90} in using a broken powerlaw to
describe $\Phi_{L}(L_{\nu})$:
\begin{equation}
  \Phi_{\rm L}(L_{\nu})=\frac{\rho_0(z)}{L_{\rm
  c}(z)}\left[\left(\frac{L_{\nu}}{L_{\rm c}(z)}\right)^{a_1} +
  \left(\frac{L_{\nu}}{L_{\rm c}(z)}\right)^{a_2}\right]^{-1}
  \label{eq:lumi}
\end{equation}
From DP90, we adopt $a_1=1.85$ and $a_2=3$.  $a_1$ is well determined
at low $z$, but at higher $z$, the flux limit of the DP90 sample
approaches $L_{\rm c}$ and an accurate determination of $a_1$ is not
possible anymore.  In fact, within the anti-hierarchical scenario for
SMBH growth \citep[e.g.][]{merloni:04} a change in slope at high
redshifts is expected, as more powerful black holes were more common
at high redshift, as indeed observed in X-ray and optically selected
AGN samples \citep{hopkins:06}.  For lack of better information, we
will assume a constant $a_1$ below.  It is reasonable and, in fact,
necessary that $\Phi_{\rm L}$ has a low-luminosity cutoff at some
minimum luminosity $L_{\rm min}$ (see \S\ref{sec:w0}).

The red-shift dependence of $L_{\rm c}$ and $\rho_0(z)$ varies with
cosmology.  DP90 adopted a $\Omega_{m}=1$, $\Omega_{\Lambda}=0$
cosmology with $H_0=50\,{\rm km\,s^{-1}\,Mpc^{-1}}$, which gave
$L_{\rm c,(1,0,50)}=6.7\times 10^{32}\,{\rm
{ergs}\,{Hz^{-1}\,s^{-1}}}\times 10^{2.35\left[1 -
\left(1+z\right)^{-1.37}\right]}$.  DP90 parameterized $\rho_0(z)$ as
$\rho_{0,(1,0,50)}=0.43 \,{Mpc^{-3}}
\times10^{\left[\sum_{i}c_{n}(0.1z)^n\right]}$ with
$c_n=\left\{-7.87,-5.74,93.06,-738.9,2248,-2399`\right\}$.

\subsection{Correcting for relativistic boosting}
Since jets are relativistic, the observed FSLF is affected by Doppler
boosting.  The Dopppler-correction of luminosity functions has been
discussed in a number of publications, most notably
\cite{urry:84,urry:91}.  Following these authors, we will neglect the
contribution from the receding jet.  The error introduced by this
approximation is small compared to the other sources of uncertainty.
The Doppler factor for a jet with Lorentz factor
$\Gamma=\sqrt{1/(1-\beta^2)}$, velocity $\beta=v/c$ and viewing angle
$\theta_{\rm LOS}$ is then given by $\delta={1}/[{\Gamma\left(1 -
\beta\cos{(\theta_{\rm LOS})}\right)]}$, with a maximum of
$\delta_{\rm max}\equiv \sqrt{(1 + \beta)/(1-\beta)}$.  The jet
luminosity for a flat spectrum source is then boosted by a factor
$\delta^2$.

Without knowledge of the underlying jet four-velocity distribution,
exact Doppler correction is impossible.  However, for sensible
distributions that show a clear peak at some $\Gamma_{\rm mean}$, it
is sufficient to approximate the distribution as a delta function,
allowing decomposition. We will assume that the velocity distribution
is well behaved in such a way.

A rest-frame (i.e., intrinsic) FSLF of the form of
eq.~(\ref{eq:lumi}), subject to Doppler boosting, will still be
observed as a broken powerlaw with the same indices, but with an
additional powerlaw regime with slope of -3/2 \citep{urry:84}.  At low
$z$, the low luminosity slope of the observed FSLF is well determined
to be steeper than $-3/2$, indicating that the $-3/2$ Doppler tail
must lie at luminosities below the DP90 flux limit.  There are
indications of a turnover to a $-3/2$ slope at luminosities below
$L_{\rm min,obs}\approx 10^{27}\,{\rm ergs\,Hz^{-1}\,s^{-1}}$ in the
low red-shift, low luminosity sample of \cite[][NFW]{nagar:05},
indicating that the {\em rest frame} FSLF becomes shallower than -3/2
below $L_{\rm min}=L_{\rm min,obs}/\delta_{\rm max}^2$.  At high $z$,
 $a_1$ is not well enough determined to draw this conclusion.

Following \cite{urry:84}, the rest frame (i.e., Doppler corrected)
FSLF must be of the form of eq.~(\ref{eq:lumi}), with the observed
break luminosity $L_{\rm c,obs}$ and the low-luminosity
cutoff/turnover $L_{\rm min,obs}$ each Doppler boosted by a factor
$(\delta_{\rm max})^2$ and the normalization corrected by a factor
\begin{equation}
  \Delta \equiv \frac{\Gamma^{2 - 2a_1}\left[(1 -
    \beta)^{3-2a_1}-1\right]}{\beta(2a_1 - 3)}.
\end{equation}

\subsection{Estimating the kinetic power normalization $W_0$}
\label{sec:w0}
To estimate the normalization $W_0$ of the radio---jet-power relation
in eq.~(\ref{eq:radiopower}), \cite{heinz:05b} and \cite{heinz:05c}
used information from three well studied radio galaxies, M87, Cygnus
A, and Perseus A, for which estimates of the kinetic power from large
scales (and kpc jet-scales in the case of M87) exist, along with
measured flat spectrum fluxes for the jet core.

\begin{figure}
\begin{center}{\resizebox{0.85\columnwidth}{!}{\includegraphics[bb=36pt
	12pt 560pt 433pt, clip=true]{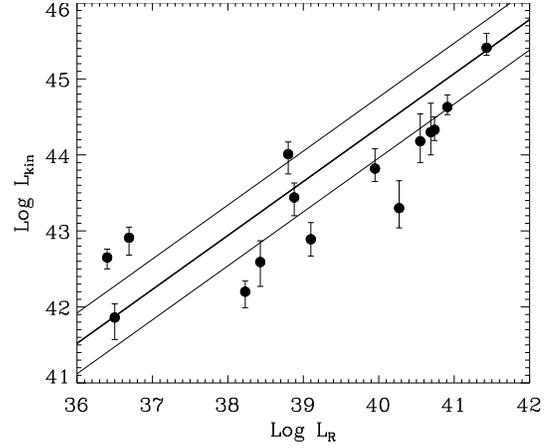}}}\end{center}
\caption{Time averaged kinetic power plotted against flat spectrum
  core radio power.  Lines show the best fit normalization (with one
  sigma uncertainties).\label{fig:sample}}
\end{figure}

Given the recent X-ray surveys of galaxy clusters with central radio
sources \citep{birzan:04,allen:06}, a more sophisticated estimate of
$W_0$ is now possible.  Taking all radio sources with robust kinetic
power estimates based on X-ray cavities and with observed nuclear flat
spectrum fluxes, we composed a sample of 15 sources. Whenever more
than one measurement of the jet kinetic power was available we have
taken the logarithmic average of the available data.
Fig.~\ref{fig:sample} plots the core flux against the kinetic power
for the 13 sources considered.  Other estimators for jet power are
available in the literature \citep[e.g.][]{willott:98}, based on total
steep spectrum radio power.  Since, unlike cavity based measurements
of $W$, these estimates are rather model dependent, we will not employ
them here.

Eq.~(\ref{eq:radiopower}) predicts that the two quantities $W_{0}$ and
$L_{\nu}$ should be related by a power-law with index $12/17$.  To
derive the constant of proportionality $W_0$ we performed a
least-squares fit to the data in Fig.~\ref{fig:sample}, fixing the
slope at $12/17$.  The best fit for $W_0$ is shown as a thick solid
line, along with two lines for the 1-sigma uncertainty derived from
the scatter in the plot.  Abitrarily fixing $L_0$ at
\begin{equation}
  L_{\rm 0,obs}\equiv 7\times 10^{29}\,{\rm ergs\,Hz^{-1}\,s^{-1}}
\end{equation}
and introducing the parameter $w_{44}\equiv W_0/10^{44}\,{\rm
ergs\,s^{-1}}$ for convenience, the best fit for $W_0$ is
\begin{equation}
  W_{0}=1^{+1.3}_{-0.6}\times 10^{44}\,{\rm ergs\,s^{-1}} \equiv
  w_{44}\times 10^{44}\,{\rm ergs\,s^{-1}}
\end{equation}

The observed core luminosity $L_{\rm 0,obs}$ must be corrected for
Doppler boosting.  For an {\em unbiased} sample of sources, the core
luminosity will, on average, be de-boosted by a factor
$\delta_{60}^2\equiv 4/[\Gamma^2(2-\beta)^2]$, corresponding to an
average viewing angle of $60^{\circ}$.  However, we cannot assess what
selection biases affect the sample of sources contributing to this
estimate \citep[e.g., X-ray sensitivity to detect jet--induced
cavities selects {\em against} beamed sources;][]{ensslin:02c}.  We
will write $L_{0}=L_{\rm 0,obs}/\delta_{60}^2$ with the implicit
understanding that $\delta_{60}$ subsumes the unknown effects of any
line-of-sight bias in the sample from Fig.~\ref{fig:sample}.

After taking all corrections into account, the un-boosted kinetic
luminosity function is
\begin{equation}
  \Phi_{W} =\frac{17\rho_{\rm 0,obs}}{12W_{\rm
  c}\Delta\delta_{\rm max}^{2-2a_{1}}}
  \left[\left(\frac{W}{W_{\rm c}}\right)^{\frac{17a_{1} -
  5}{12}}+\left(\frac{W}{W_{\rm c}}\right)^{\frac{17a_{2} -
  5}{12}}\right]^{-1}
\label{eq:kineticluminosityfunction}
\end{equation}
where we defined the critical power $W_{\rm c}$ as
\begin{equation}
  W_{\rm c}\equiv W_{0}\left(\frac{L_{\rm c,obs}\delta_{\rm
      60}^2}{L_{\rm 0,obs}\delta_{\rm max}^2}\right)^{\frac{12}{17}}
\end{equation}

$\Phi_{W}$ has a low luminosity slope of $\Phi(W) \propto W^{(5 -
17a_{1})/12}$.  This implies that the total integrated kinetic power
is dominated by the lowest power sources for values of $a_1$ steeper
than $a_{1} > 29/17 \sim 1.7$.  Given that DP90 measured $a_1 \approx
1.85$, the luminosity function {\em must} have a break or cutoff at
some $L_{\rm min}$ somewhere below the flux limit of DP90 (possibly
given by the value of $L_{\rm min,obs}$ indicated by NFW), since
otherwise the total kinetic luminosity would diverge.

\section{Discussion} 
\label{sec:discussion}
\begin{figure}
\resizebox{\columnwidth}{!}{\includegraphics{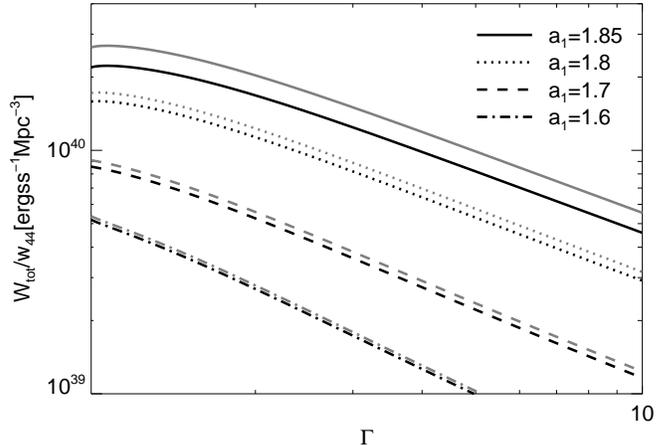}}
\caption{Integrated jet power density for different parameter
  combinations at $z=0$, as a function of mean jet Lorentz factor
  $\Gamma_{\rm mean}$.  Solid lines: $a_1=1.85$, dotted lines: $a_1 =
  1.8$, dashed lines: $a_1=1.7$, dash-dotted lines: $a_1=1.6$; black
  lines: $L_{\rm min}=2.5\times 10^{27}\,{\rm ergs\,Hz^{-1}\,s^{-1}}$
  (DP90), grey lines: $L_{\rm min}=10^{27}\,{\rm
  ergs\,Hz^{-1}\,s^{-1}}$ (NFW).\label{fig:power}}
\end{figure}

The total power released by jets from all sources under the FSLF
(i.e., all black holes except those contributing to the observed
steep--spectrum luminosity function, abbreviated as SSLF), per
comoving Mpc$^3$, is simply the first moment of the kinetic luminosity
function:
\begin{equation}
  W_{\rm tot}(z)  =  \frac{17 \rho_{\rm 0,obs}(z)\,W_{\rm
  c}}{12\,\Delta\,\delta_{\rm max}^{2-2a_{1}}}
  \int_{x_0}^{\infty}\frac{dx}{x^{\frac{17}{12}(a_1 - 1)} +
  x^{\frac{17}{12}(a_2 - 1)}} 
  \label{eq:kineticluminosity}
\end{equation}
where $x_0(z) \equiv \left(L_{\rm min,obs}(z)/L_{\rm
  c,obs}(z)\right)^{\frac{12}{17}}$.  For $L_{\rm min} \approx
  10^{27}\,{\rm ergs\,Hz^{-1}\,s^{-1}}$ (NFW), this yields
\begin{eqnarray}
  W_{\rm tot}(z=0) & \approx & 2.8 \times 10^{40}{\rm ergs\,s^{-1}
  Mpc^{-3}}\,\mathcal{D}\,w_{44}\\ \mathcal{D} & \equiv &
  \left(\frac{\delta_{60}}{\delta_{\rm
  max}}\right)^{\frac{24}{17}}\frac{1}{\Delta\delta_{\rm
  max}^{2-2a_1}}
\end{eqnarray}
This is the estimated total power released per Mpc$^3$ by flat
spectrum jets today.  The main sources of uncertainty in $W_{\rm tot}$
are $W_0$, $a_1$, $\Gamma_{\rm mean}$, and, for large values of $a_1$,
the estimate of $L_{\rm min}$.  Fig.~\ref{fig:power} shows $W_{\rm
tot}/w_{44}$ as a function of $\Gamma_{\rm mean}$ for different values
of $L_{\rm min}$ and $a_1$.

To put this value of $W_{\rm tot}$ in context, it is useful to compare
it to the average stellar luminosity density $L_{\star} \approx
2\times10^{41}\,{\rm ergs\,s^{-1}\,Mpc^{-3}}$ \citep{ellis:96} and to
the current supernova power $W_{\rm SN} \approx 10^{39}\,{\rm
ergs\,s^{-1}\,Mpc^{-3}}$ \citep{madau:98b}, which is well below the
integrated jet power.

The estimate of $W_{\rm tot}$ is dominated by low luminosity sources
(which is why $L_{\rm min}$ is critical for $W_{\rm tot}$).  This
suggests that AGN make a much bigger contribution to feedback in
regular galaxies than commonly assumed.  It also implies that low
luminosity AGN dominate the global kinetic energy output of black
holes in a quasi-steady state, rather than short, energetic bursts of
individual black holes.  This picture agrees well with the concept of
slow, ``effervescent'' feedback envisioned to be responsible for AGN
heating in galaxy clusters \citep{begelman:01,churazov:02}.

In order to derive the total energy density released by jets, we can
integrate $W_{\rm tot}$ from eq.~(\ref{eq:kineticluminosity})
over redshift, taking the proper cosmological corrections into
account.  We will use an upper limit of $z_{\rm max}=5$ for the
redshift integral, but the results are not sensitive to the exact
value of $z_{\rm max}$.  Taking the most conservative approach by
using the observed radio flux limit from DP90 as a solid {\em upper}
limit on the low-luminosity cutoff $L_{\rm min}$, we find a lower
limit of
\begin{equation}
  e_{\rm tot} = \int_{0}^{\infty}dz \frac{dt}{dz} W_{\rm tot} > 1.5\times
  10^{57}{\rm ergs\,Mpc^{-3}}\,\mathcal{D}\,w_{44}
  \label{eq:etot}
\end{equation}
on the total integrated jet energy density.  A more realistic
assumption would be to adopt the low--redshift value of $L_{\rm min}$
from NFW and assume the same redshift evolution for $L_{\rm min}$
as that of $L_{\rm c}$.  This gives an estimated value of
\begin{equation}
  e_{\rm tot} \approx 1.7\times 10^{58}{\rm ergs\,
  Mpc^{-3}}\,\mathcal{D}\,w_{44}
  \label{eq:etot2}
\end{equation}
Keeping $L_{\rm min}$ fixed at the $z=0$ value from NFW at all
redshifts provides a robust upper limit of $e_{\rm tot} < 2.5\times
10^{58}\,{\rm ergs\,\,Mpc^{-3}}\mathcal{D}w_{44}$.

Comparing $e_{\rm tot}$ to the mean cosmic black hole mass density of
$\rho_{\rm BH} \approx 3.3\times 10^{5}\,M_{\odot}\,{\rm
Mpc^{-3}}\,h_{75}^2$ \citep{yu:02} finally yields the average
conversion efficiency of accreted rest mass to jet power for
supermassive black holes:
\begin{equation}
  \epsilon_{\rm jet} \equiv \frac{e_{\rm tot}}{\rho_{\rm BH}c^2}
  \approx 3\%\,\mathcal{D}\,w_{44}
  \label{eq:epsilon}
\end{equation}
for the same assuptions that went into eq.~(\ref{eq:etot2}).  Assuming
that $L_{\rm min}(z)$ is smaller than the survey flux limit and larger
or equal to $L_{\rm min,NFW}$. gives limits of
$0.25\%\,\mathcal{D}\,w_{44} < \epsilon_{\rm jet} <
4.3\%\,\mathcal{D}\,w_{44}$.  This estimate of $\epsilon_{\rm jet}$ is
broadly consistent with previous best-guess estimates of the jet
conversion efficiency, typically believed to be of the order of 1\% -
10\%.

Note, however, that the estimate of $\rho_{\rm BH}$ from \cite{yu:02}
includes mass accreted in all phases of black hole growth.  Black
holes grow predominantly through radiatively efficient accretion: The
observed amount of X-ray background radiation is equivalent to about
10\% of the total black hole rest mass energy measured today, implying
that, for typical radiative efficiencies of order 10\%, {\em most} of
the accreted mass must have contributed to the production of the X-ray
background in a radiatively efficient mode \citep{soltan:82}.
Radiatively efficient accretion flows are typically radio quiet (i.e.,
inefficient at producing jets).  Thus, the conversion efficiency
during low-luminosity accretion phases must be significantly {\em
higher} than the average $\epsilon$ implied by eq.~(\ref{eq:epsilon}).
Given that about 10\% of AGN are radio loud, the average black hole
accumulates a fraction of $f_{m} = 90\% f_{90}$ of its mass during
radio quiet, radiatively efficient accretion.  Thus, the jet
conversion efficiency $\epsilon_{\rm jet}$ during radio loud phases
must be at least a factor of $(1-f_{m})^{-1} \sim 10$ larger than
shown in eq.~(\ref{eq:epsilon}), of order 30\%.

Efficiencies of several tens of percent are not implausible, however:
If black hole spin is important in jet launching \citep{blandford:77}
and if black holes accrete large amounts of angular momentum
\citep[i.e., are close to maximally rotating for a significant
fraction of their life, as suggested by recent merger-tree
models][]{volonteri:05}, they can liberate up to 30\% of the accreted
rest mass energy by black hole spin extraction alone, implying
$\epsilon_{\rm jet}$ of up to 42\%.

Finally, we will briefly discuss the possible contribution of
steep--spectrum sources to $\epsilon_{\rm jet}$.  By definition, the
FSLF contains all black holes except those included in the SSLF (which
are dominated by optically thin synchrotron emission).  The scaling
relation from eq.~(\ref{eq:radiopower}) does not hold for
steep--spectrum sources.  We can, however, derive an upper limit on
the contribution from steep spectrum sources.  Assuming a typical
optically thin synchrotron spectral index of 0.65, any underlying flat
spectrum component would have to fall below 20\% of the observed 5GHz
luminosity, otherwise the source would become too flat to qualify as a
steep--spectrum source.  Using the SSLF from DP90, the same redshift
integral that yielded eq.~(\ref{eq:etot}) provides an upper limit on
the contribution from the flat spectrum sources underneath the SSLF of
$\epsilon_{\rm steep} < 6\%\,\tilde{\mathcal{D}}\,w_{44}$, where
$\tilde{\mathcal{D}}$ is the Doppler correction for the steeper
spectral index. Since the SSLF is shallow at low luminosities,
$\epsilon_{\rm steep}$ is dominated by sources around $L_{\rm c}$ and
not affect by the uncertainty in $L_{\rm min}$.

\section{Summary}
\label{sec:summary}
Starting from the relation between kinetic jet power and flat spectrum
core radio luminosity, we derived the kinetic luminosity function of
flat spectrum radio sources.  We found that the kinetic luminosity
density is dominated by the lowest luminosity sources, indicating that
constant, low level effervescent type heating is important in black
hole feedback.  Integrating the kinetic luminosity density over
redshift and comparing it to the estimate of the current black hole
mass density showed that the efficiency of jet production by black
holes is of the order of a few percent and smaller than 10\%.
However, since most of the power comes from low luminosity sources,
which are not believed to contriubte much to the total mass accretion
of the black hole, the efficiency of jet production during low
luminosity, jet-driven phases must be significantly higher.

\end{document}